\documentclass[usenatbib]{mn2e}
\usepackage{times,graphics,amssymb,epsfig,subfigure,color}

\newcommand{\ber}{\begin{eqnarray}}
\newcommand{\eer}{\end{eqnarray}}

\def\apj{ApJ}

\def\labequn #1{\label{eq:#1}}
\def\labfig #1{\label{fig:#1}}
\def\labsecn #1{\label{sec:#1}}
\def\labsubsecn #1{\label{subsecn:#1}}
\def\labsubsubsecn #1{\label{subsubsecn:#1}}
\def\labtablem #1{\label{tab:#1}}

\def\equn #1{Equation~\ref{eq:#1}}

\def\fig #1{Figure~\ref{fig:#1}}
\def\dfig #1#2{Figures~{\ref{fig:#1}}~and~{\ref{fig:#2}}}
\def\secn #1{Section~\ref{sec:#1}}

\def\subsecn #1{Section~\ref{subsecn:#1}}
\def\dsubsecn #1#2{Sections~{\ref{subsecn:#1}}~and~{\ref{subsecn:#2}}}
\def\subsubsecn #1{Section~\ref{subsubsecn:#1}}
\def\tablem #1{Table~\ref{tab:#1}}

\def\etal{et al.\ }

\def\unit #1{\,{\rm #1}}

\def\kev{\unit{keV}}

\title[Thermodynamic stability of BHB winds]
{The effects of thermodynamic stability on wind properties in different low mass black hole binary states}
\author[Chakravorty \etal]
{
Susmita Chakravorty$^{1,2}$
Julia Lee$^{1,2}$,
Joseph Neilsen$^{3,4}$ \\
\footnotesize \it $^{1}$Harvard University, Department of Astronomy, schakravorty@head.cfa.harvard.edu; \\ 
\it $^{2}$Harvard-Smithsonian Center for Astrophysics, 60 Garden Street, Cambridge, MA 02138, USA \\
\it $^{3}$Einstein Fellow, Boston University, 725 Commonwealth Avenue, Boston, MA 02215, USA \\
\it $^{4}$MIT Kavli Institute for Astrophysics and Space Research, Cambridge, MA 02139, USA.
}

\begin{document}
\maketitle


\begin{abstract}

We present a systematic theory-motivated study of the thermodynamic stability
condition as an explanation for the observed accretion disk wind signatures in
different states of low mass black hole binaries (BHB). The variability in
observed ions is conventionally explained either by variations in the driving
mechanisms or the changes in the ionizing flux or due to density effects,
whilst thermodynamic stability considerations have been largely ignored. It
would appear that the observability of particular ions in different BHB states
can be accounted for through simple thermodynamic considerations in the static
limit. Our calculations predict that in the disk dominated soft thermal and
intermediate states, the wind should be thermodynamically stable and hence
observable. On the other hand, in the powerlaw dominated spectrally hard state
the wind is found to be thermodynamically unstable for a certain range of $3.55
\le \log \xi \le 4.20$. In the spectrally hard state, a large number of the
He-like and H-like ions (including e.g. Fe\, {\sc xxv}, Ar\, {\sc xviii} and
S\, {\sc xv} have peak ion fractions in the unstable ionization parameter
($\xi$) range, making these ions undetectable. Our theoretical predictions have
clear corroboration in the literature reporting differences in wind ion
observability as the BHBs transition through the accretion states \citep{lee02,
miller08, neilsen09, blum10, ponti12, neilsen12a}.  While this effect may not
be the only one responsible for the observed gradient in the wind properties as
a function of the accretion state in BHBs, it is clear that its inclusion in
the calculations is crucial to understanding the link between the environment
of the compact object and its accretion processes.

\end{abstract}




\section{Introduction}
\labsecn{sec:introduction}

Most stellar mass black hole binaries (BHBs) show common behaviour patterns
centered around a few states of accretion. The different accretion states are
signified by different spectral energy distributions (among other things)
having varying degree of contribution from the accretion disk and the
non-thermal powerlaw components. In addition, since \textit{Chandra} and
XMM-Newton, there has been increased interest in winds from the accretion disk,
as a result of detections of blueshifted absorption lines of varying velocities
and temperatures, seen in high resolution X-ray spectra. In order to get a
consolidated picture of these systems, it is necessary to understand the
relation between the accretion states of the BHBs and the properties of the
accretion disk winds. 

The \textit{Chandra} HETGS (High Energy Transmission Grating Spectrometer,
\citealt{canizares05}) sensitive 0.5-7.5 keV energy range is rich in atomic
transitions. Yet often absorption lines from only H- and He-like Fe are
detected (e.g \citealt{lee02}, \citealt{neilsen09} for GRS 1915+105,
\citealt{miller04} for GX 339-4, \citealt{miller06} for H1743-322 and
\citealt{king12} for IGR J17091-3624). On the other hand in some rare cases a
range of ions is seen, from O through Fe (e.g. \citealt{ueda09} for GRS
1915+105, \citealt{miller08, kallman09} for GRO J1655--40).  This suggests that
the wind properties, e.g. temperature and density, may vary depending on the
source and/or the accretion state.  In this paper we restrict our discussion to
low mass BHBs to ensure that wind signatures originate from the accretion disk
rather than from the stellar companion (as in the case of high mass X-ray
binaries).  However, we note that accretion disk winds are detected in neutron
star binary systems as well (e.g \citealt{brandt00}, \citealt{schulz02a} for
Cir X-1, \citealt{ueda04} for GX 13+1, \citealt{reynolds10} for 1A0535+262 and
\citealt{miller11} for IGR J17480-2446) and that some high mass X-ray binaries
(where the companion to the compact object is a high mass star) exhibit stellar
winds diagnosed through emission (e.g. Vela X-1, \citealt{schulz02b}; LMC X-4,
\citealt{neilsen09b}) or absorption (e.g. Cyg X-1, \citealt{hanke09}).

X-ray studies of BHBs show that winds are not present in all states.
\citet{neilsen09} demonstrated that the equivalent width of the Fe XXVI
absorption line is anti-correlated with the fractional hard power law
contribution to the spectrum in GRS 1915+105: the softer the state, the more
prominent the absorption lines; see also \citet{miller08, blum10}.
Recently a confirmation of this finding came from the \citet{ponti12}
compilation of wind results (by the aforementioned and other authors) verifying
the state dependent nature of accretion disk winds in general. 

As an explanation for the observed changes in these winds, authors have
commonly invoked differences in photoionizing flux \citep[e.g.][for
H1743-322]{miller12}. However, since the properties of winds are also a result
of their driving mechanisms \citep{lee02, ueda09, ueda10, neilsen11,
neilsen12a}, observed wind variations may also indicate changes in these
driving mechanisms. For example, a well known \textit{Chandra} observation of
GRO J1655--40 \citep{miller06, miller08, kallman09}, showing a rich absorption
line spectrum with ions from OVIII to NiXXVI. \citet{miller06, miller08} argue
in favour of the magnetic driving mechanism for the wind. 
But a \textit{Chandra} observation from 3 weeks earlier, for the same object,
shows only Fe\, {\sc xxvi} absorption \citep{neilsen12a}. The changes cannot be
explained by changes in the photoionizing flux alone, and \citet{neilsen12a}
suggest that variable thermal pressure and magnetic fields may both be
important in driving long-term changes in this wind.

Interestingly while changes in the photoionizing flux, in the driving
mechanisms and in the wind density are frequently invoked to explain the
observed wind behaviour, the thermodynamic stability of these outflows its
relevance to the detectability of particular ions in BHBs have never been
discussed in detail. Although quite commonly used for assessing winds in active
galactic nuclei \citep{krolik81, reynolds95, hess97, chakravorty09,
chakravorty12, lee13}, thermodynamic stability arguments have not been used
much to explain observed BHB wind behaviour. (However, see
\citealt{jimenez-garate02} for a specific use of the stability curves for
determining accretion disk atmosphere properties for neutron star binaries.)
Using thermal equilibrium curves, in this paper we test the importance of
thermodynamic stability for the observable properties of winds in different
accretion states of BHBs. We are particularly interested in testing if the
prevalence of the winds in the softer states \citep[as noted
by][etc.]{neilsen09, ponti12} can be explained using thermodynamic stability
arguments.


\section{The Spectral Energy Distribution}
\labsecn{sec:SED}

We begin our considerations by adopting appropriate spectral energy
distributions (SEDs) for different BHB states.  The SED of BHBs usually
contains, to varying degrees, (1) a thermal component conventionally modeled
with a multi-temperature blackbody originating from the inner accretion disk
and often showing a characteristic temperature ($T_{in}$) near 1 keV and (2) a
non thermal power-law component with a photon spectrum $N(E) \propto
E^{-\Gamma}$ \citep{remillard06}. The thermal state is dominated by the
radiation from the inner accretion disk contributing more than 75\% of the 2-20
keV flux (e.g.  dashed red line on top panel of \fig{fig:Seds}). In contrast
the hard power law state (a.k.a. hard state; solid black line in
\fig{fig:Seds}) is dominated by a flat power law component ($\Gamma\sim1.8$)
that contributes more than 80\% of the 2-20 keV flux.  For any given BHB, the
accretion disk usually appears to be fainter and cooler in this hard power law
state than it is in the thermal state.

The radiation from a thin accretion disk may be modeled as the sum of local
blackbodies emitted at different radii. 
A simple prescription for the radial distribution of the temperature is:
\begin{equation} 
T(R) = 6.3 \times 10^5 \left(\frac{\dot{m}} {\dot{m}_{\rm Edd}} \right)^{\frac{1}{4}} \left(\frac {M_{\rm{BH}}} {10^8M_{\odot}} \right)^{-\frac{1}{4}} \left(\frac{R}{R_s}\right)^{-\frac{3}{4}}\rm{K} 
\labequn{eqn:DbbTemp}
\end{equation}
\citep{peterson97,frank02} where $\dot{m}$ is the accretion rate of the central
black hole of mass $M_{\rm{BH}}$, $\dot{m}_{\rm Edd}$ is its Eddington accretion
rate and $R_s = 2GM_{\rm{BH}}/c^2$ is the Schwarzschild radius ($G$ is the
gravitational constant and $c$ is the velocity light). To describe the
radiation from the accretion disk $f_{disk}(\nu)$, we use the
\citet{zimmerman05} model {\tt ezdiskbb} (from
XSPEC,\footnote{http://heasarc.gsfc.nasa.gov/docs/xanadu/xspec/}
\citealt{arnaud96}). The model {\tt ezdiskbb} imposes the physical boundary
condition that the viscous torque should be zero at the inner edge of the disk
at radius $R_{in}$. The model is parametrised by  
\begin{equation}
T_{max} = 0.488 f T_{in} = 0.488 f T(R_{in})
\labequn{eqn:EzdbbTemp}
\end{equation}
and 
\begin{equation} 
A_{ezdbb} = \left(1/f^4\right) \left\{ \frac {R_{in}/\rm{km}} {D/(10\,\,\rm{kpc})} \right\}^2 \cos\theta 
\labequn{eqn:EzdbbNorm}
\end{equation}
where \equn{eqn:DbbTemp} is used to calculate $T(R_{in})$; $D$ is the distance
of the source from us, $\theta$ is the angle that our line-of-sight makes to the
normal to the disc plane; $f$ is the hardening factor which accounts for the
modification of the optically thick disk emission from a pure blackbody. We use
the established value $f=1.7$ \citep{shimura95}

\begin{figure}
\begin{center}
\includegraphics[scale = 1, width = 9 cm, trim = 0 130 0 0, clip, angle = 0]{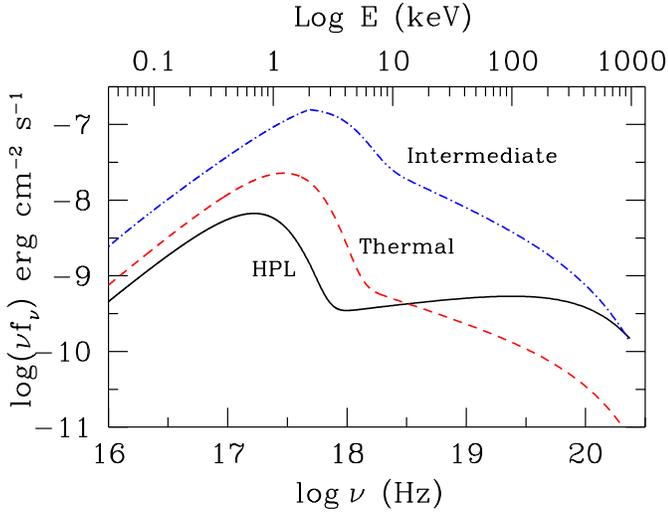}
\caption{The SEDs for the three different accretion states of a
black hole of mass $7 M_{\odot}$ at a distance of 5 kpc from us. The two SED
components (disk and powerlaw) are added following the schemes described in
\citet{remillard06}. See \secn{sec:SED} for the details.} 
\labfig{fig:Seds}
\end{center}
\end{figure}

We add a power-law component to the disk spectrum using
\begin{equation}
f(\nu) = f_{disk} (\nu) + [A_{pl} \nu^{-\alpha}] \exp{^{-\frac{\nu}{\nu_{max}}}}
\labequn{eqn:FullSed}
\end{equation}
to account for the full BHB SED, where $\alpha = \Gamma-1$ is the spectral
index of the powerlaw. 

In their review, \citet{remillard06} used the different SEDs of the 1996-97
outburst of the BHB GRO J1655--40 \citep[hereafter
GROJ1655,][]{sobczak99,orosz97} to demonstrate the SEDs of a typical BHB in its
different states. Following their prescription, we define three different
fiducial SEDs for the three states of a $7 M_{\odot}$ black hole at a distance
of 5 kpc from us. (Note that the the distance of the black hole from us does
not have any consequence on the thermodynamic calculations presented in this
paper for the absorbing gas which is very close to the black hole. To determine
the photoionization state of the gas, it is sufficient to know the shape of the
SED and the ionization parameter. Thus the exact value of the distance is not
important.)  We also find in \subsubsecn{subsubsec:BhPars} that a variation in
$4 M_{\odot} \le M_{\rm{BH}} \le 15 M_{\odot}$ does not strongly affect our
overall conclusions.) Subsequent results presented in this paper will be based
on the following SED definitions we adopt for the three typical low mass BHB
states.
\begin{enumerate}
\item[$\bullet$] {\bf Thermal state} (\fig{fig:Seds} dashed red curve): To generate
the disk spectrum, we choose $\left( \dot{m}/\dot{m}_{\rm Edd}, R_{in}/R_s
\right) = (0.2, 6.0)$ resulting in $T_{max}$ ($T_{in}$) = 0.48 (0.58) \kev. The
powerlaw has $\Gamma = 2.5$ and $A_{pl}$ is chosen in such a way that the 2-20
keV disk flux contribution $f_d = 0.8$. 
\item[$\bullet$] {\bf Intermediate state} (\fig{fig:Seds} dotted blue curve): The
disk spectrum is generated with $\left( \dot{m}/\dot{m}_{\rm Edd}, R_{in}/R_s
\right) = (0.4, 3.0)$ so that $T_{max}$ ($T_{in}$) = 0.97 (1.17) \kev. For the
powerlaw, $\Gamma = 2.7$ and $f_d = 0.5$.
\item[$\bullet$] {\bf Hard powerlaw (hereafter HPL) state} (\fig{fig:Seds} solid
black curve): With $\left( \dot{m}/\dot{m}_{\rm Edd}, R_{in}/R_s \right) =
(0.1, 10.0)$ we generate a cooler disk with $T_{max}$ ($T_{in}$) = 0.28 (0.33)
\kev. The powerlaw is dominant in this state with $\Gamma = 1.8$ and $f_d =
0.2$. 
\end{enumerate}
For each of the SEDs defined above, we use a high energy exponential cut-off
(\equn{eqn:FullSed}) to insert a break in the power law at $E_{b}=100$ keV
(\fig{fig:Seds}). See \subsubsecn{subsubsec:PlCutOff} for a discussion of the
effects of varying $E_{b}$. (In soft states $E_b$ can be less than 20 keV, and we accommodate this in our considerations of thermodynamic stability for soft states; see \subsubsecn{subsubsec:PlCutOff} for details.)

Note that even though we base BHB state SEDs on GRO J1655
\citep[following][]{remillard06}, our results are intended to be generic to
stellar mass black holes (in so far as the 1996-97 outburst of GRO J1655 was
representative of a typical BHB outburst). We use these SEDs as reasonable
representations of the ionizing spectra for  different BHB states, and we use
similar representative parameters for the wind properties (e.g.
density, column density, and metallicity). In subsequent sections, we make
quantitative tests of the influence of these choices, and find that our results
are robust to changes in the SED and wind properties (including $M, \,\,
\dot{m} \,\, \rm{and} \,\, R_{in}$ in \subsubsecn{subsubsec:BhPars}; $f_d$ and
$\Gamma$ and in \subsubsecn{subsubsec:PlCutOff}; $E_{b}$ in
\subsubsecn{subsubsec:PlCutOff} and wind density in \subsecn{subsec:density}.

\begin{figure}
\begin{center}
\includegraphics[scale = 1, width = 9 cm, angle = 0]{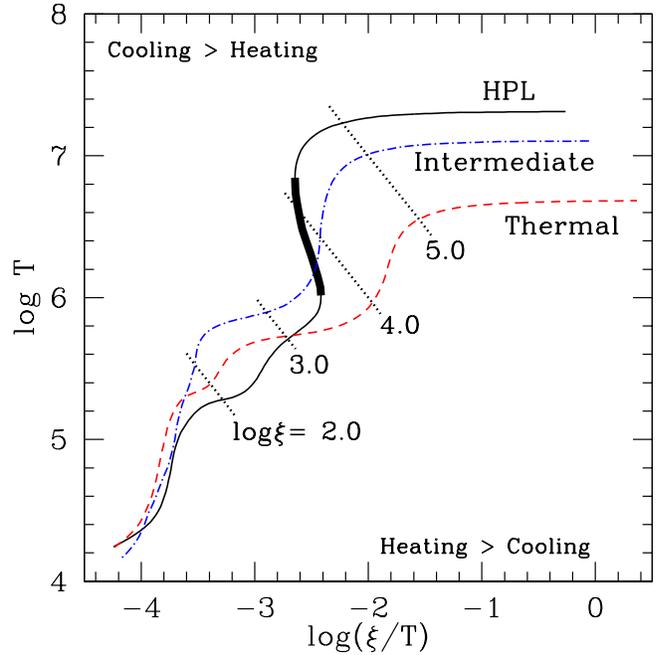}
\caption{Stability curves for photoionised Solar metallicity gas with density
$n_{\rm H} = 10^{14} \rm{cm^{-3}}$ and column density $N_{\rm H} = 10^{23}
\rm{cm^{-2}}$.  The different curves correspond to the three different ionizing
continuum for the three different accretion states of the black hole, described
in \fig{fig:Seds} and \secn{sec:SED}. The dotted black lines cutting across the
stability curves mark regions of constant $\xi$, also demonstrating the range
of $\xi$ spanned by the stability curves. The highlighted region (thick black
line) of the stability curve for the HPL state shows the range of
thermodynamically unstable phases. The other two stability curves are stable
throughout.}  
\labfig{fig:Scurves}
\end{center}
\end{figure}

\section{Photoionization calculations for thermodynamic equilibrium}
\labsecn{sec:Calculations}

In this Section we use the aforementioned (\fig{fig:Seds}; \secn{sec:SED})
ionizing SEDs for the different BHB states to calculate the expected
temperature, pressure (\subsecn{subsec:Scurves}) of the outflowing gas, and the
ionization fractions (\subsecn{subsec:IF}) of the various ions in the gas. This
will allow us to assess the thermodynamic stability of winds as a means of
explaining the presence/absence of observable ionic species in different BHB
states.

\begin{table*}
\begin{center}
\begin{tabular}{cc ccc ccc cc}
\hline
\raisebox{-1.5ex}[0cm][0cm]{$\log \xi$} && \multicolumn{2}{c}{HPL} && \multicolumn{2}{c}{Thermal} && \multicolumn{2}{c}{Intermediate} \\
&& Heating & Cooling && Heating & Cooling && Heating & Cooling \\
\hline 
&& & && \multicolumn{2}{c}{$n_{\rm H} = 10^{6} \rm{cm^{-3}}$} &&  & \\
&& $^a$HI, n=2 (25.7) & $^{b}$H iso. of H (42.1) && HI, n=2 (24.4) & H iso. of H (35.6) && HI, n=2 (21.1) & H iso. of H (23.3) \\
&& HeI (18.2) & FeII recomb.$^{c}$ (20.4) && OI (15.8) & FeII recomb. (22.0) && OI (18.1) & FeII recomb. (23.2) \\
&& OII (12.9) & MgII 2795.5 $\AA$ (6.0) && HeI (11.5) & MgII 2795.5 $\AA$ (5.8) && FeII (10.1) & FeII recomb. (9.9) \\ \cline{3-10}
&& & && \multicolumn{2}{c}{$n_{\rm H} = 10^{10} \rm{cm^{-3}}$} &&  & \\
&& OII (30.2) & H iso. of H (33.7) && OII (28.0) & H iso. of H (49.9) && OII (20.7) & H iso. of H (55.0) \\
0.0 && HeI (27.4) & MgII 2795.5 $\AA$ (15.7) && HeI (15.3) & Free-free (13.5) && HeI (10.7) & Free-free (15.6) \\
&& CII (7.5) & MgII 2802.7$\AA$ (10.6) && NeII (6.6) & MgII 2795.5 $\AA$ (11.5) && FeII (9.1) & MgII 2795.5 $\AA$ (8.3) \\ \cline{3-10}
&& & && \multicolumn{2}{c}{$n_{\rm H} = 10^{14} \rm{cm^{-3}}$} &&  & \\
&& O II (25.5) & H recomb. (33.0) && O II (27.4) & H recomb. (33.0) &&  O II (25.2) & H recomb. (32.9) \\
&& H I (15.5) & Free-free (26.4) && H I (11.8) & Free-free (27.2) &&  H I (12.8) & Free-free (24.2) \\
&& He II (15.2) & H iso. of H (19.6) && Fe III (11.3) & H iso. of H (19.0) && Fe III (11.5) & H iso. of H (23.8) \\
\hline
&& & && \multicolumn{2}{c}{$n_{\rm H} = 10^{6} \rm{cm^{-3}}$} &&  & \\
&& HeII (49.6) & CIV 1548$\AA$ (13.0) && HeII (33.7) & CIV 1548$\AA$ (17.6) && HI (28.6) & H iso. of H (46.9)\\
&& OIV (10.2) & OIII 5007$\AA$ (8.1) && OIV (11.1) & OIII 5007$\AA$ (9.1) && OIII (17.0) & OIII 5007$\AA$ (12.7) \\
&& OIII (7.7) & CIV 1551$\AA$ (6.5) && OIII (10.3) & CIV 1551$\AA$ (8.8) && HeII (10.5) & CIII 1910$\AA$ (8.6) \\ \cline{3-10}
&& & && \multicolumn{2}{c}{$n_{\rm H} = 10^{10} \rm{cm^{-3}}$} &&  & \\
&& HeII (40.2) & CIV 1548$\AA$ (32.3) && HeII (25.4) & CIV 1548$\AA$ (30.9) && OII (25.7) & Si III 1892$\AA$ (16.5)\\
1.0 && OIV (11.0) & CIV 1551$\AA$ (17.2) && OIII (14.7) & CIV 1551$\AA$ (16.7) && SiIII (6.7) & CII 1335$\AA$ (14.5) \\
&& OIII (10.6) & CIII 977$\AA$ (4.4) && OIV (10.6) & CIII 977$\AA$ (7.3) && FeIII (6.6) & H iso. of H (12.0) \\ \cline{3-10}
&& & && \multicolumn{2}{c}{$n_{\rm H} = 10^{14} \rm{cm^{-3}}$} &&  & \\
&& O V (21.9) & Free-free (20.2) && O V (22.1) & Free-free (15.3) && O IV (16.2) & Free-free (12.0) \\
&&  O IV (18.2) & H recomb. (13.4) && O IV (15.8) & H recomb. (9.7) && O III (15.4) & H recomb. (8.5) \\
&& Ne V (6.3) & H iso. of H (10.0) && Ne V (7.6) & H iso. of H (6.5) && Ne IV (6.7) & H iso. of He (6.7) \\
\hline
&& & && \multicolumn{2}{c}{$n_{\rm H} = 10^{6} \rm{cm^{-3}}$} &&  & \\
&& HeII (17.2) & Free-free (19.8) && OVIII (16.1) & Free-free (25.4) && OVIII (13.9) & Free-free (23.1) \\
&& OVIII (11.5) & SiXII 499$\AA$ (5.6) && HeII (15.4) & H iso. of He (14.3) && FeXVIII (9.6) & H iso. of He (18.3) \\
&& FeXIV (7.1) & H recomb. (5.6) && FeXVIII (7.7) & H recomb. (10.3) && FeXIX (8.3) & H recomb. (5.4) \\ \cline{3-10}
&& & && \multicolumn{2}{c}{$n_{\rm H} = 10^{10} \rm{cm^{-3}}$} &&  & \\
&& HeII (16.0) & Free-free (20.4) && OVIII (16.4) & Free-free (25.8) && OVIII (13.9) & Free-free (23.1) \\
2.0 && OVIII (11.6) & FeXV 284.2$\AA$ (6.1) && HeII (14.2) & H iso. of He (13.3) && FeXVIII (9.6) & H iso. of He (17.7) \\
&& FeXIV (7.5) & SiXII 499$\AA$ (6.0) && FeXVIII (8.0) & H recomb. (6.5) && FeXIX (8.1) & H recomb. (5.5) \\ \cline{3-10}
&& & && \multicolumn{2}{c}{$n_{\rm H} = 10^{14} \rm{cm^{-3}}$} &&  & \\
&& O VIII (15.8) & Free-free (22.1) && O VIII (20.0) & Free-free (25.2) && O VIII (17.7) & Free-free (24.6) \\
&& He II (7.8) & H recomb. (9.4) && Fe XVIII (9.1) & H recomb. (9.5) && Fe XIX (12.1) & H iso. of He (8.8) \\
&& Fe UTA$^{d}$ (7.4) & He recomb. (6.7) && Fe XIX (9.0) & H iso. of He (7.8) && Fe XVIII (8.9) & H recomb. (7.7) \\
\hline
&& & && \multicolumn{2}{c}{$n_{\rm H} = 10^{14} \rm{cm^{-3}}$} &&  & \\
&& Fe XX (14.9) & Compton (43.0) && Fe XXII (14.4) & Free-free (39.7) && Fe XXIII (14.1) & Free-free (46.0) \\
3.0 && Compton (13.3) & H recomb. (9.3) && O VIII (14.4) & H recomb. (8.6) && O VIII (14.0) & Fe XXIV - 192 \AA (7.8) \\
&& Fe XIX (11.5) & He recomb. (9.2) && Fe XXI (13.8) & He recomb. (8.5) && Fe XXII (9.8) & He recomb. (7.8) \\ 
\hline
&& & && \multicolumn{2}{c}{$n_{\rm H} = 10^{14} \rm{cm^{-3}}$} &&  & \\
&& Compton (79.2) & Free-free (69.1) && Compton (35.7) & Free-free (60.4) && Compton (62.5) & Free-free (69.8) \\
4.0 && Fe XXV (6.3) & Compton (16.0) && O VIII (11.8) & He recomb. (10.0) && Fe XXV (10.5) & Compton (13.0) \\
&& Fe XXIV (2.1) & He recomb. (3.8) && Fe XXIV (11.1) & H recomb. (8.7) && O VIII (4.8) & He recomb. (5.4) \\
\hline
&& & && \multicolumn{2}{c}{$n_{\rm H} = 10^{14} \rm{cm^{-3}}$} &&  & \\
&& Compton (98.3) & Compton (82.0) && Compton (88.1) & Compton (66.0) && Compton (9.7) & Compton (77.4) \\
5.0 &&  Fe XXVI (1.0) & Free-free (16.6) && Fe XXV (3.8) & Free-free (28.1) && Fe XXVI (1.8) & Free-free (20.2) \\
&& Fe XXV (0.1) & Fe XXV recomb. (0.3) && Fe XXVI (3.1) & He recomb. (1.8) && O VIII (0.2) & Fe XXV recomb. (0.5) \\ 
%
%
\hline
\end{tabular} \\
$^a$ HI, n=2 ---$>$ photoionization from the n=2 level of neutral Hydrogen \\
$^b$ H iso. ---$>$ Hydrogen iso-sequence \\
$^c$ recomb. ---$>$ recombination \\
$^{d}$ UTA ---$>$ Unresolved transition array (2p-3d inner-shell absorption by iron M-shell ions)
\caption{The heating and cooling agents as a function of $\log \xi$ for the
HPL, thermal and intermediate stability curves. For $\log \xi \le 2.0$ we have
listed the agents for three different densities $n_{\rm H} = 10^{6}, 10^{10}
\,\, \rm{and} \,\, 10^{14} \rm{cm^{-3}}$, whereas for higher ionization
parameters (which are insensitive to density variations; see
\subsecn{subsec:density}) we have given the list for only $n_{\rm H} = 10^{14}
\rm{cm^{-3}}$. The numbers in the parentheses give the fractional heating or
cooling contributed (in percentage) by the particular agent. We have listed
only the top three contributing agents in each case. When a given ion is listed
as the heating agents it is its photoionization which is responsible for
heating the gas.} 
\labtablem{table1}
\end{center}
\end{table*}

\subsection{Stability curves}
\labsubsecn{subsec:Scurves}

Thermodynamic stability of photoionized gas can be studied effectively using
the thermal equilibrium curve of the temperature ($T$) versus the pressure
($\xi/T$) of the gas \citep{krolik81, reynolds95, hess97, chakravorty08,
chakravorty09, chakravorty12, lee13}; here $\xi = L/n_{\rm H}r^2$ is the
ionization parameter. Along the stability curve, heating and cooling are
balanced everywhere, but gas in thermal equilibrium is only thermodynamically
stable where the slope of the curve is positive. The thermodynamic
stability of the photoionized gas is determined by the balance of the major
heating and cooling mechanisms along the curves (see \tablem{table1} for an
example list of such agents, varying with $\xi$). We expect that if the
photoionized gas (as diagnosed through observed absorption lines) is detected,
its ionization parameter is such that the gas is in thermal equilibrium. Hence
the $\xi/T-T$ point of such gas should lie on the stable parts of the stability
curve.

We model the photoionized absorber as an optically thin, plane parallel
slab of Solar metallicity \citep{allendeprieto01, allendeprieto02, holweger01,
grevesse98} gas with density $n_{\rm H} = 10^{14} \rm{cm^{-3}}$ and column
density $N_{\rm H} = 10^{23} \rm{cm^{-2}}$ (but see \subsecn{subsec:density}
for a discussion on effects caused by density variation). Using each of the
three fiducial BHB SEDs as the ionizing continuum for this gas we generate
three different stability curves (\fig{fig:Scurves}) using version C08.00 of
CLOUDY\footnotemark \citep{ferland98}.
\footnotetext{URL: http://www.nublado.org/ }
%

\begin{figure}
\begin{center}
\includegraphics[scale = 1, width = 9 cm, trim = 0 130 0 70, clip, angle = 0]{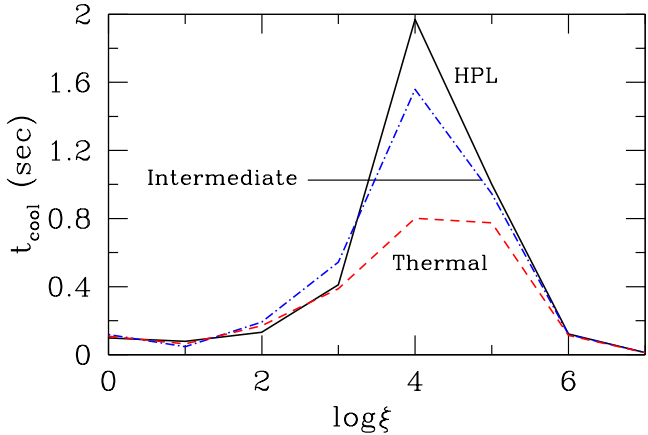}
\caption{The thermal cooling time scale as a function of $\log \xi$ for all
three SEDs, namely, HPL, thermal and intermediate. The density used for the
calculations is $n_{\rm H} = 10^{14} \rm{cm^{-3}}$ and column density $N_{\rm
H} = 10^{23} \rm{cm^{-2}}$, for solar metallicity gas.} 
\labfig{fig:tcool}
\end{center}
\end{figure}

The stability curves span a wide range of $\xi$ values (\fig{fig:Scurves}
dotted black lines cutting across the curves), for most of which all the
curves are stable. However, we find that during the HPL state
(\fig{fig:Scurves} solid black), hot plasma with $3.55 \le \log \xi \le 4.20$
is thermodynamically unstable. The ions associated with such gas are therefore
unobservable. In contrast, during the intermediate and thermal states, we find
that the ionized gas is thermodynamically stable throughout $0 \le \log \xi \le
6$, and thus a range of ions from low to high Z (atomic number) elements can be
detected in these states. All else being equal (e.g. luminosity, density of the
gas, and its distance from the central source), detections of ionized winds
should be much more common in thermal and intermediate states. Our predictions
of the thermodynamic properties of winds in BHB are consistent with the
observations that winds are detected in softer, more thermal states and are
rarely detected in spectrally hard states (here represented by HPL, e.g.
\citealt{miller08, neilsen09, ponti12}).

In \tablem{table1} we compare the major heating and cooling agents affecting
the different BHB state stability curves in an effort to understand what causes
the gas ionized by the HPL SED to become unstable at $3.55 \le \log \xi \le
4.20$. As can be seen, at $\log \xi \sim 4$, Compton heating is far more
dominant for the HPL SED ionized gas (79\%) than for the thermal SED ionized
gas. This affects the temperature, which is higher, due to Comptonization in
the former case, and renders the HPL equilibrium curve unstable in the
aforementioned $\xi$ range. 

Note that for the calculations in \dsubsecn{subsec:Scurves}{subsec:IF} we have
used a gas with constant density $10^{14} \rm{cm^{-3}}$. However, we tested the
effect of varying the densities between $10^5 - 10^{14} \rm{cm^{-3}}$ in
\subsecn{subsec:density} and find that the stability curves are {\it
insensitive} to density variations for $\log \xi \gtrsim 3.0$, leaving the
above mentioned results for the range $3.55 \le \log \xi \le 4.20$ for the HPL
state, unchanged. 

\subsubsection{Thermal time scales}
\labsubsubsecn{constraint}

The photoionization calculations performed in this paper are conducted in the
static limit. In assessing the relevance of our conclusions for observed ionic
differences between BHB states, we calculate the thermal cooling time scale
$t_{cool}$ (defined as the time in which the gas looses half of the heat gained
by it; \fig{fig:tcool}) of the gas in the static limit. While calculating
$t_{cool}$ CLOUDY considers all the heating and cooling processes associated
with static gas (some of which are detailed in \tablem{table1}). The static gas
assumption is reasonable if $t_{cool}$ is less than the adiabatic cooling time
scale $t_{ad}$ ($\sim \left[n_{\rm H}kT(a/u + 2u/r)\right]^{-1}$, where $a$,
$u$ and $r$ are the acceleration, wind velocity, and radius respectively, $k$
being the Boltzmann constant). Calculation of the adiabatic or dynamical time
scale is beyond the scope of this paper (but see \citealt{begelman83, krolik83,
chelouche05, luketic10} for example discussions related to the dynamics of
outflowing gas in AGN and X-ray binaries). We do not make any assumptions about
the dynamics or the launching mechanisms of the gas.  The purpose of this paper
is to study the thermodynamics of the photoionized gas, for which we have
assumed that the static limit requirements are met.


\subsection{Ion Fractions}
\labsubsecn{subsec:IF}

\begin{figure*}
\begin{center}
\includegraphics[scale = 1, width = 17 cm, trim = 0 40 0 20, clip, angle = 0]{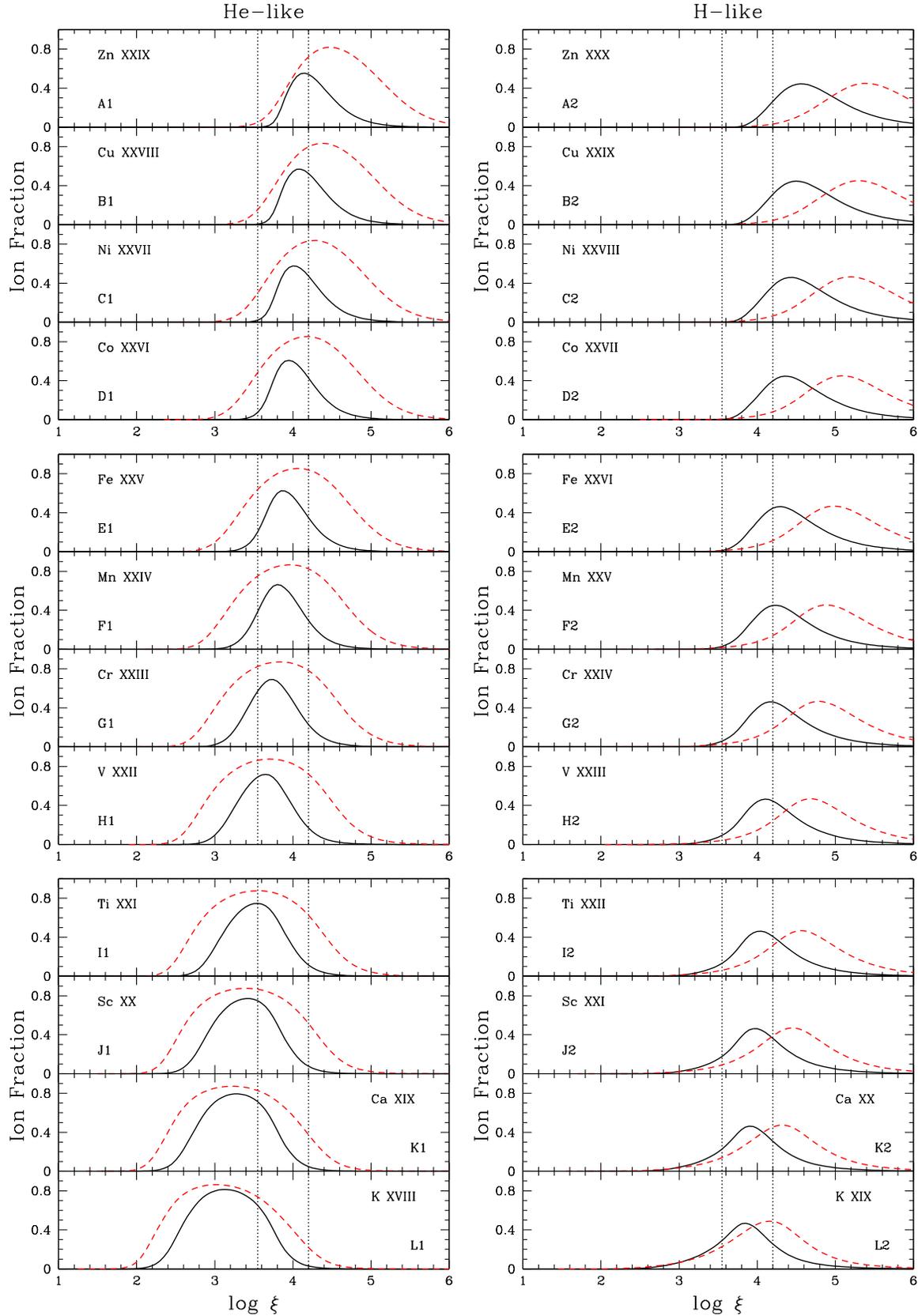}
\caption{The ion fraction distributions as a function of $\log \xi$ for the HPL
(solid black line) and the thermal state (dashed red line) of a BHB. The
vertical black dotted lines denote the range $3.55 \le \log \xi \le 4.20$ over
which the absorbing gas is {\it thermodynamically unstable if} ionized by the
HPL state SED. Note that this unstable $\xi$ range does not apply to the thermal
state ion fraction distributions, because the the thermal state stability
curves are thermodynamically stable for all values of $\xi$.} 
\labfig{fig:IF}
\end{center}
\end{figure*}

\setcounter{figure}{3}
\begin{figure*}
\begin{center}
\includegraphics[scale = 1.0, width = 17 cm, trim = 0 160 0 20, clip, angle = 0]{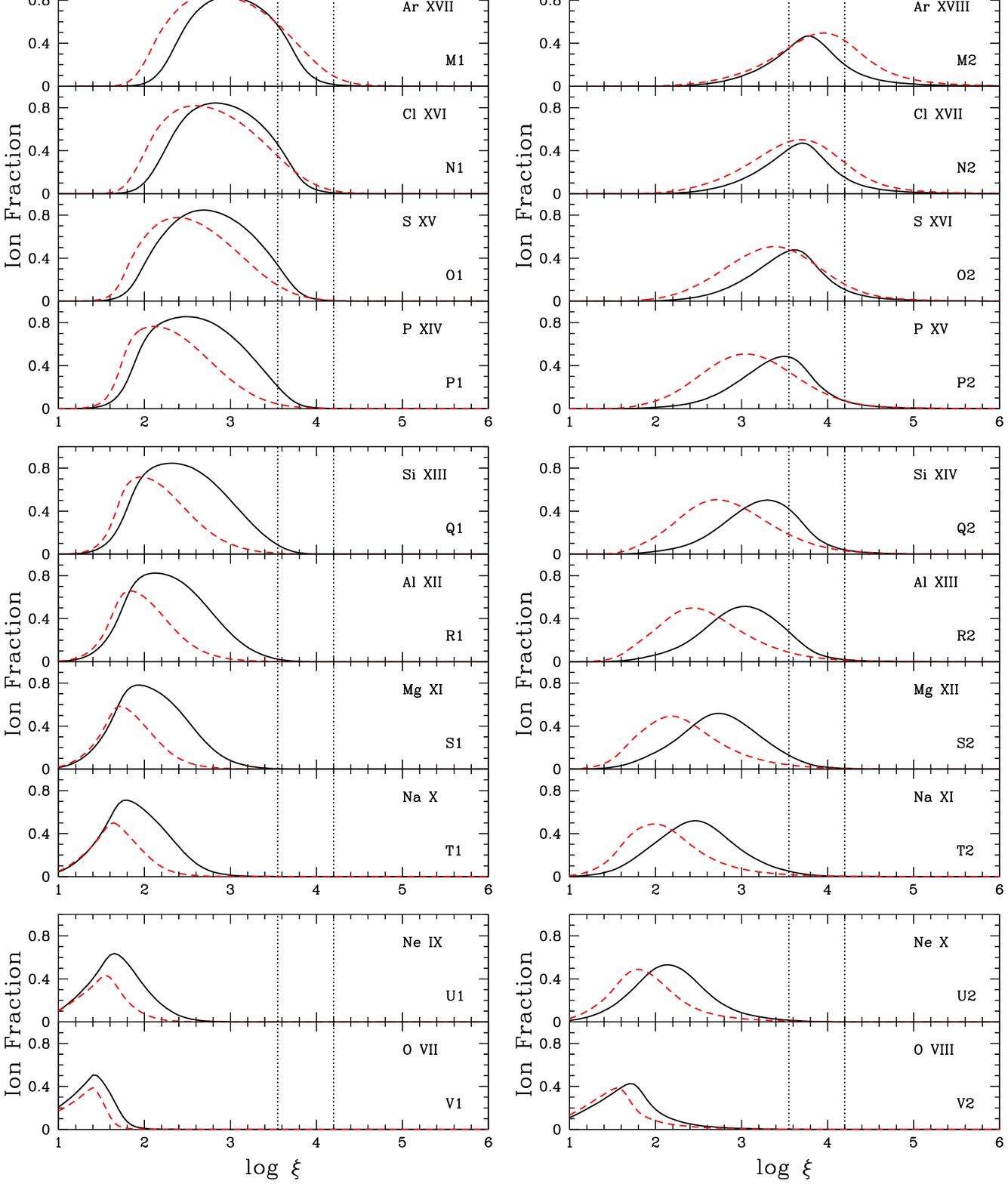}
\caption{Continued for ions of elements with lower atomic number.} 
\end{center}
\end{figure*}

In this section we further investigate the range $3.55 \le \log \xi \le 4.20$
in terms of ion fractions, to assess thermodynamic instability effects on ion
observability. In \fig{fig:IF} we plot the $\log \xi$ distribution of the ion
fractions of the He-like and H-like ions of all the elements from
zinc (atomic no.  Z = 30) to oxygen (Z = 8). Vertical lines mark
thermodynamically unstable range of $\log \xi$ based on our HPL ionizing SED.

For a given ion if a significant part of the HPL state ion fraction
distribution falls within the unstable $3.55 \le \log \xi \le 4.20$ range,
absorption lines of that ion would not be detected. Thus, during the HPL state
a number of important He- and H-like species (including Fe\,{\sc xxv}) are
essentially `shrouded' by the thermodynamically unstable $\xi$ range.  Purely
from the thermodynamic point of view, these species are not expected to be
visible during spectrally hard states. It is notable, however, that the peak
ion fraction for Fe\,{\sc xxvi} (and higher Z H-like ions) falls outside the
unstable $\xi$ range. In other words, if there is a wind in a spectrally hard
state with sufficiently high ionization ($\log \xi>4.2$), it could still be
detected in Fe\,{\sc xxvi} absorption.

The unstable $3.55 \le \log \xi \le 4.20$ range does not apply to the
thermal/intermediate states, since there is no thermal instability. As can be
seen from \dfig{fig:Scurves}{fig:IF}, all ions should be observable from a
thermodynamic stability point of view. 

These results agree nicely with observational results on the long-term
variability of winds. As discussed above, winds are commonly detected through
Fe\,{\sc xxv} and Fe\,{\sc xxvi} absorption during spectrally soft states, and
rarely detected during harder states. There are a small number of exceptions,
however. For example, in a 2005 \textit{Chandra} HETGS observation of GROJ1655
during a spectrally hard state, a single Fe\,{\sc xxvi} absorption line was
seen \citep{miller08, neilsen12a}. \citet{neilsen12a} explicitly note the
absence of Ar\,{\sc xviii}, S\,{\sc xv}, and Si\,{\sc xiv} in that state. By
our calculations, the absence of these ions is not surprising, given that the
required $\xi$ for the gas producing these ions would have fallen within the
thermodynamically unstable $\xi$ range.


\begin{figure}
\begin{center}
\includegraphics[scale = 0.5, width = 9 cm, trim = 50 0 0 0, angle = 0]{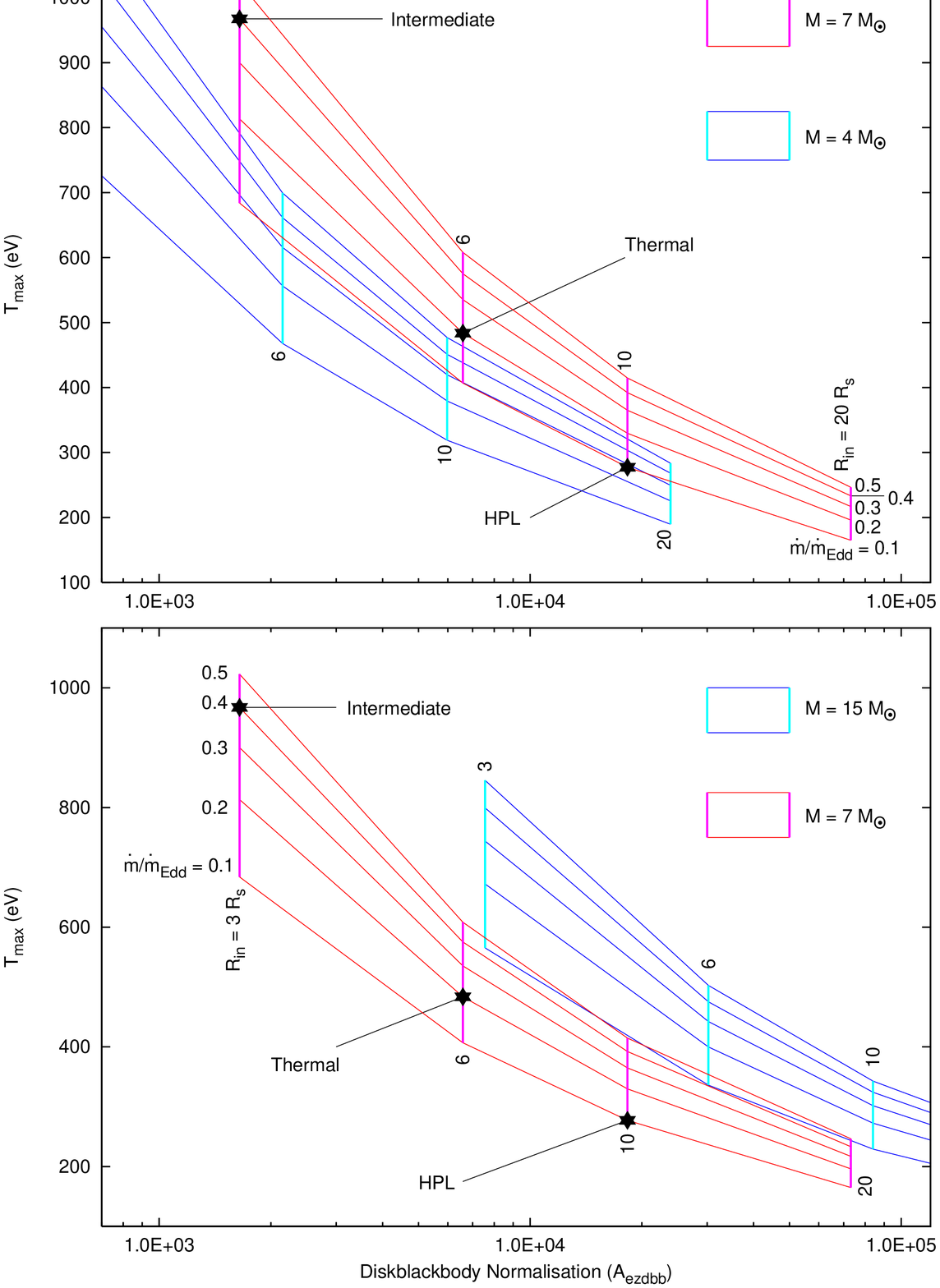}
\caption{The $T_{max}$ - $A_{ezdbb}$ distribution as a function of the various
black hole parameters, $M$, $\dot{m}/\dot{m}_{\rm Edd}$ and $R_{in}/R_s$. The
fiducial values of the $T_{max}$ and $A_{ezdbb}$ used to generate the SEDs in
this paper, are marked as solid black stars on the grid corresponding to
$M_{\rm{BH}} = 7M_{\odot}$ (red-and-magenta). In the top panel we compare the
standard $M_{\rm{BH}} = 7M_{\odot}$ distribution to that of a lower $M_{\rm{BH}} =
4M_{\odot}$ black hole, and in the bottom panel with that of an higher $M_{\rm{BH}}
= 15M_{\odot}$ one.} 
\labfig{ParamGrids}
\end{center}
\end{figure}


\section{Discussion}
\labsecn{sec:discussion}

The fiducial SEDs discussed in \secn{sec:SED} and shown in \fig{fig:Seds},
although generic representations of the respective states, were generated using
certain SED parameters. In this section we investigate the effects of varying
SED parameters -  $M, \,\, \dot{m} \,\, \rm{and} \,\, R_{in}$
(\subsubsecn{subsubsec:BhPars}), $f_d$ and $\Gamma$
(\subsubsecn{subsubsec:PlPars}) and powerlaw break $E_b$
(\subsubsecn{subsubsec:PlCutOff}) on thermodynamic stability results. We find
that physically and observationally reasonable variations of these parameters
do not alter the conclusions of our thermodynamic calculations
(\secn{sec:Calculations}). The same is true for variations in the wind density
(\subsecn{subsec:density})

\subsection{Variations in the SED parameters}

\subsubsection{Variations in $M, \,\, \dot{m} \,\, \rm{and} \,\, R_{in}$}
\labsubsubsecn{subsubsec:BhPars}


A plausible range of black hole masses lie between $M \sim 4 M_{\odot}$ (e.g.
XTE J1650--500 \citealt{orosz04}) and $M \sim 15 M_{\odot}$ (e.g see
\citealt{greiner01} for GRS 1915-105, \citealt{orosz07} for M 33 X-7 and
\citealt{orosz11} for Cygnus X-1). As discussed in \secn{sec:SED}, we choose a
black hole of mass $M_{\rm{BH}} = 7 M_{\odot}$ (based on GROJ1655) to represent a
generic BHB. We further chose particular sets of $\dot{m}/\dot{m}_{\rm Edd}$
and $R_{in}/R_s$ that correspond to values of $T_{max}$ and $A_{ezdbb}$
typically observed in outbursts of BHBs. 

Given the relationship between temperature, disk radius, accretion rate, and
black hole mass (\equn{eqn:DbbTemp}), how much are the SEDs affected by
variations in $M_{\rm{BH}}$? In \fig{ParamGrids}, we have compared the relationship
between $T_{max}$ and $A_{ezdbb}$ for a $7 M_{\odot}$ black hole to that of a
$4 M_{\odot}$ and a $15 M_{\odot}$ black hole. The fiducial values for our
three SEDs are marked as solid black stars. The top panel of \fig{ParamGrids}
shows that for reasonable accretion rates and inner disk radii, our thermal/HPL
SEDs could apply to a $4 M_{\odot}$ black hole; the intermediate SED requires
higher accretion rates ($\dot{m} > 0.5 \dot{m}_{\rm Edd}$), but not unheard of
in BHBs. The $15 M_{\odot}$ black hole can also match the fiducial values of
$T_{max}$ with reasonable values of $\dot{m}/\dot{m}_{\rm Edd}$ and
$R_{in}/R_s$, although the implied disk normalizations are a factor of $\sim 5$
larger. Thus we believe that our SEDs can be considered broadly representative
of BHBs at a range of masses, accretion rates, and disk radii.


\subsubsection{Variations in disk fraction $f_{disk}$ and $\Gamma$}
\labsubsubsecn{subsubsec:PlPars}

As discussed, we have followed the scheme of \citet{remillard06} to pick
reasonable combinations for  accretion disk and power law components (see
\secn{sec:SED}). Since our stability curves are dependent on overall spectral
shape, it is imperative to test if our results (derived from the stability
curves) are robust to variations in $f_d$ and $\Gamma$

\citet{remillard06} define the thermal state as having $f_d > 0.75$. Keeping
our $\Gamma$ constant at 2.5, we calculated new stability curves for $f_d (0.7,
0.8, 0.9)$, and found no changes in the thermodynamic stability of the gas.
Since \citet{remillard06} do not state a particular range for $\Gamma$, we vary
$1.4 \le \Gamma \le 3.0$ (in steps of 0.2) keeping $f_d = 0.8$ constant.  For
$\Gamma < 2.0$ the curve becomes unstable in the range $3.84\le \log \xi \le
4.24$.  However, a literature survey reveals no instances of $\Gamma < 2$ in the
thermal state. Hence we conclude that during the thermal state, winds should be
thermodynamically stable for all values of $\xi$.

In addition \citet{remillard06} define the spectrally hard state (which we call
the HPL state) as having $f_d < 0.2$ and $1.4 \le \Gamma \le 2.1$. Varying $f_d
(0.1, 0.2)$ while keeping $\Gamma = 1.8$ constant and then varying $1.4 \le
\Gamma \le 2.2$ (in steps of 0.2) while keeping $f_d = 0.2$ constant we find
that the curve retains the unstable $\xi$ range discussed in
\secn{sec:Calculations}.

Thus the qualitative results presented in \secn{sec:Calculations} hold for the
entire range of the parameters of X-ray emission defined by \citet{remillard06}
for the BHB states.


\begin{figure*}
\begin{center}
\includegraphics[scale = 1.0, height = 19 cm, trim = 230 0 130 30, angle = 90]{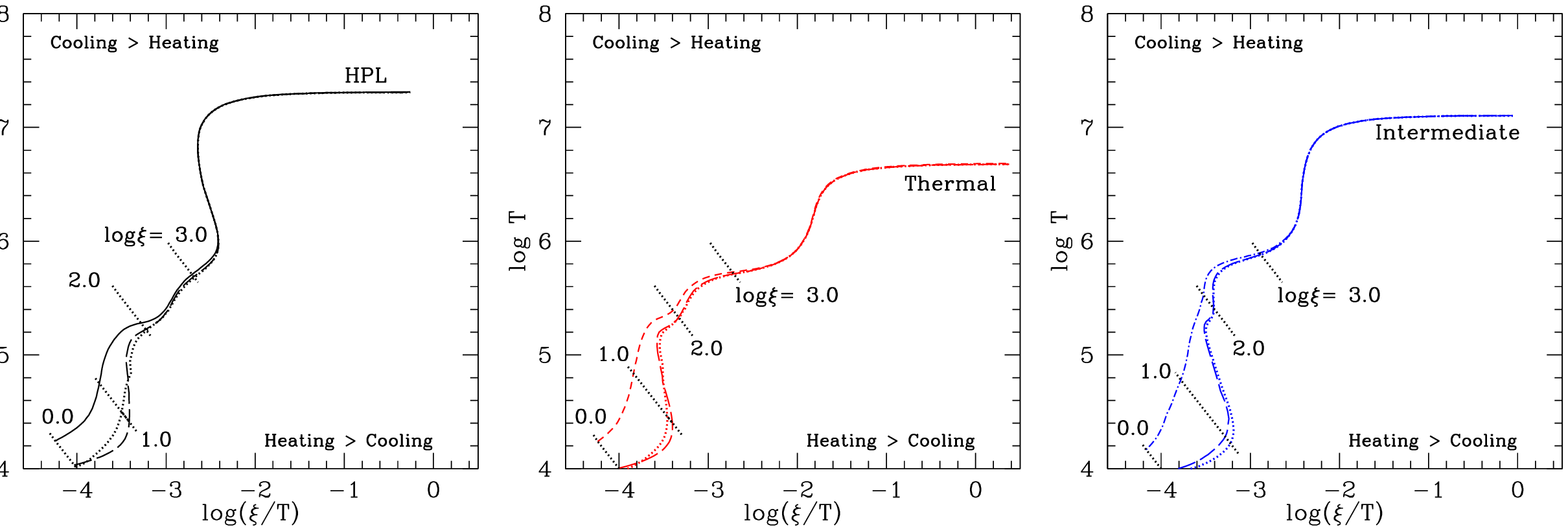}
\caption{The effects of density variation on the stability curves are shown for
the three BHB states (HPL in the left panel, thermal in the middle panel and
intermediate in the right panel). Other than the fiducial density value of
$n_{\rm H} = 10^{14} \rm{cm^{-3}}$ (for which we maintain the same line styles
as in \fig{fig:Scurves} - dashed for thermal, dashed-and-dotted for
intermediate and solid for HPL), in each panel we consider two additional
densities $n_{\rm H} = 10^{10} \rm{cm^{-3}}$ (long-dashed) and $n_{\rm H} =
10^6 \rm{cm^{-3}}$ (dotted). For each SED the stability curves are identical
for $\log \xi \ge 3.0$, showing that density does not affect the heating and
cooling at these ionization parameters. However, for lower $\xi$, the constant
$\xi$, diagonal dotted black line segments are maintained to show the range of
ionization parameter where density effects are important.}  
\labfig{DensityVariation}
\end{center}
\end{figure*}

\subsubsection{Variations in the powerlaw break energy $E_b$}
\labsubsubsecn{subsubsec:PlCutOff}

For all the SEDs discussed in the previous sections, the parameters for the
exponential cut-off in the powerlaw (\equn{eqn:FullSed}) were chosen such that
the powerlaw had a break at $E_b \sim 100 \kev$. Observations of spectrally
hard (HPL) states of BHBs show powerlaw breaks at $E_b \sim 100 \kev$, but for
thermal states the break is usually observed at much lower energies $E_b \sim
20 - 30 \kev$. It is believed that this has to do with the presence or absence
of jets (and associated particle acceleration processes) in these states.
Regardless of the origin of changes in the break energy, though, the effect on
thermodynamic conditions is worth investigating, so we calculated new stability
curves for $E_b = 20, 40, \,\,\rm{and} \,\, 80 \kev$, both for the thermal and
HPL states. 

Based on our analysis, the stability curves are insensitive to the variation in
$E_b$ for $\log {\rm T} < 5.7$ (K) or $\log \xi < 3.0$. At higher temperatures,
the stability curves become more smooth and the plasma actually becomes more
stable as the break energy decreases from 100 keV to 20 keV. This does not
affect our conclusions for the thermal state, since the plasma was already
stable at all $\xi$.

However, one consequence of variations in $E_b$ is that during the HPL state,
the aforementioned range $3.55 \le \log \xi \le 4.20$ becomes thermodynamically
stable (see \secn{sec:Calculations}) as $E_b$ is decreased to 20 keV. 
The theoretical reason for this effect is easy to understand, where one recalls
the discussion on heating and cooling agents in \subsecn{subsec:Scurves}.
According to \tablem{table1}, at high ionization parameters ($\log \xi \sim
4.0)$, the temperature of the gas ionized by HPL SED is determined by Compton
heating (79\% contribution) when $E_b = 100 \kev$.  However, if $E_b \sim 20
\kev$, then there is a significant decrease in the number of high energy
photons thus reducing the extent of Comptonization (Comptonization contributes
only 49\% to heating) . Hence the temperature of the gas decreases resulting in
a thermodynamically stable distribution of temperature and pressure.  Although
theoretically noteworthy, this is not a relevant result because the jet
dominated HPL state is not likely to have such low break in the powerlaw.
Thus, variation in the powerlaw break $E_b$ also does not affect the results in
\secn{sec:Calculations}, unless there are reasons to consider that the HPL
state continuum has a low powerlaw break at $\sim 20 \kev$.
 


\subsection{Variations in wind density}
\labsubsecn{subsec:density}

We vary the particle density of the absorbing gas between $n_{\rm H} = 10^5 -
10^{14} \rm{cm^{-3}}$ and generate stability curves (\fig{DensityVariation})
for the HPL state (left panel), the thermal state (middle panel) and the
intermediate state (right panel) SEDS, to determine if the results presented in
\secn{sec:Calculations} are susceptible to density effects. As can be seen from
the figure, density variations do not have any effect on the stability curves
for $\log \xi \gtrsim 3.0$ and the effect is small until $\log \xi < 2.0$. For
lower $\log \xi$ the stability curves are seen to be thermodynamically unstable
if the wind has density $n_{\rm H} \lesssim 10^{10} \rm{cm^{-3}}$.  Because the
vast majority of plasma in BHBs can be described by ionization parameters $\log
\xi \sim 4$ (with the exception of GX 339-4, $\xi \sim 70$;
\citealt{miller04}), our conclusions for the thermodynamic stability (or lack
of it) of winds are not affected by gas density variations.


Even though we primarily see high $\xi (\log \xi \gtrsim 4)$ gas in BHBs, we
still consider density effects at low $\xi (\log \xi \lesssim 2.0)$ to improve
theoretical understanding. \fig{DensityVariation} shows that the greatest
difference in the gas temperature occurs at $\log \xi \sim 1.0$ in comparing
plasma with $n_{\rm H} \le 10^{10} \rm{cm^{-3}}$  to one with $n_{\rm H} =
10^{14} \rm{cm^{-3}}$. Referring back to \tablem{table1} we find the detailed
heating and cooling agents corresponding to the different stability curves.
Specifically, comparing the results for $\log \xi = 1.0$ in \tablem{table1},
the heating and cooling agents and their fractional contributions are similar
for $n_{\rm H} = 10^6 \,\, \rm{and} \,\, 10^{10} \rm{cm^{-3}}$ but differ from
that of $n_{\rm H} = 10^{14} \rm{cm^{-3}}$. An explanation is that - line
cooling is dominant for $n_{\rm H} \le 10^{10} \rm{cm^{-3}}$, whereas continuum
processes like free-free cooling, recombination cooling and radiation through
emission of iso-sequences of Hydrogen and Helium are dominant cooling
mechanisms for $n_{\rm H} = 10^{14} \rm{cm^{-3}}$ resulting in higher
temperatures in the later case.


\section{Conclusions}
\labsecn{conclusions}

We have constructed fiducial SEDs for the thermal, intermediate and HPL states
of the BHBs, following the prescription of \citet{remillard06} for a typical $7
M_{\odot}$ black hole, based on GROJ1655. Using these SEDs we generated
thermodynamic stability curves for each state to arrive at the following
conclusions:
\begin{enumerate}
\item[$\bullet$] In the thermal and intermediate state all phases of the wind
is {\it thermodynamically stable} (\S \subsecn{subsec:Scurves}). However, the
ionization parameter range $3.55 \le \log \xi \le 4.20$ is {\it
thermodynamically unstable} for winds in the HPL state. We found that a large
number of the He-like ions (21 $<$ Z $<$ 30, Ti through Cu) and H-like ions (15
$<$ Z $<$ 25, S through Mn) have peak ion fractions in the unstable ionization
parameter range for the HPL state, making the lines from these ions potentially
unobservable (\S \subsecn{subsec:IF}).
\item[$\bullet$] Our findings are well corroborated in the observational
literature, since there appears to be a gradient in wind properties with BHB
state. In accord with their thermodynamic stability (and lack of it for
spectrally harder HPL states), winds are predominantly observed in intermediate
and soft/thermal states and have weak or absent lines from Fe\,{\sc xxv} and
low-Z elements in the HPL states \citep{lee02, miller08, neilsen09, blum10,
ponti12, neilsen12a}. In the HPL state, while the ions with peak ion fractions
in the unstable $\xi$ range may be unobservable, Fe\,{\sc xxvi} may
remain detectable at high ionization parameters ($\log\xi>4.2$) in such states
(based on our calculations in \subsecn{subsec:IF} and consistent with
observations noted by \citealt{neilsen12a}).
\item[$\bullet$] In general, a range of ionizations are thermodynamically
stable (except for $3.55 \le \log \xi \le 4.20$ for HPL state) in all the
states we have studied here. Yet the observed absorbers are usually consistent
with a single high ionization parameter. This may indicate a genuine absence of
plasma at lower ionization. On the other hand, for many other sources
intervening cold absorption from the source or the ISM may be the reason.
\item[$\bullet$] There have been suggestions for magnetically driven winds in
e.g GROJ1655 \citep{miller08, neilsen12a}. 
The pressure term in the temperature - pressure distributions (stability
curves) considered in this paper deals with only the thermodynamic gas pressure
on the photoionised medium (through $\xi/T$).  Magnetic pressure in not
included in such calculations. It would be interesting to incorporate the
magnetic pressure term, which would change the stability conditions from what
we have shown in this paper. Given a reasonable description of the magnetic
field in the photoionized medium, codes like CLOUDY are capable of such
calculations. In our future publications we shall formulate this phase space as
a diagnostic tool for photoionized gas which are under the influence of strong
magnetic fields.
\end{enumerate}


\section*{acknowledgments}

We thank Gary Ferland for making CLOUDY publicly available and for providing
helpful tips. We acknowledge the generous support of the \textit{Chandra}
theory grant TM3-14004X that is administered by the Smithsonian Astrophysical
Observatory.




\end{document}